\documentclass{ws-procs9x6}

\RequirePackage{xspace}
\usepackage{graphicx}% Include figure files
\usepackage{dcolumn}% Align table columns on decimal point
\usepackage{bm}
\usepackage{relsize}

\def\babar{\mbox{\slshape B\kern-0.1em{\smaller A}\kern-0.1em
    B\kern-0.1em{\smaller A\kern-0.2em R}}}
\def\Kbar{\kern 0.2em\overline{\kern -0.2em K}{}\xspace}
\def\Kzb{\Kbar^0\xspace}
\def\Bbar    {\kern 0.18em\overline{\kern -0.18em B}{}\xspace}
\def\Bzb     {\ensuremath{\Bbar^0}\xspace}

\begin{document}

\title{Charmless Hadronic $B$ Decays at \babar{} %\footnote{\uppercase{T}his work is supported by etc, etc.}
}

\author{J. BIESIADA %\footnote{\uppercase{W}ork partially supported by grant 2-4570.5 of the \uppercase{S}wiss \uppercase{N}ational \uppercase{S}cience \uppercase{F}oundation.}
}

\address{Department of Physics \\
Princeton University \\
Princeton, NJ 08544\\ 
E-mail: biesiada@princeton.edu}

\maketitle

\abstracts{
We present recent results on charmless hadronic $B$ decays using data 
collected by the \babar{} detector at the PEP-II asymmetric-energy $e^+e^-$ 
collider at the Stanford Linear Accelerator Center.  We report measurements 
of branching fractions and charge asymmetries in several charmless 
two-body, three-body, and quasi-two-body decay modes.  We also report
measurements of polarization in charmless B decays to exclusive final states
with two vector mesons.}

\section{Introduction}

Charmless hadronic $B$ decays are an important probe of the standard model (SM).  They can be used to test the Cabibbo-Kobayashi-Maskawa (CKM) mechanism of flavor mixing and $CP$ violation, with sensitivity to the three angles $\alpha$, $\beta$, and $\gamma$ of the Unitarity Triangle for $B$ decays.  Charmless processes are usually dominated by $b \to u$ tree amplitudes and ``penguin'' decays mediated by $b\to s$ and $b\to d$ processes involving a virtual loop with the emission of a gluon.  These transitions are suppressed by CKM factors in the SM, with branching fractions at the $10^{-6}-10^{-5}$ level.  Contributions from supersymmetric particles or other physical effects beyond the SM could induce observable deviations from SM predictions in the measured branching fractions and $CP$ asymmetries.\cite{Penguins}

In these proceedings, I summarize the most recent measurements for this class of decays at the \babar{} experiment at SLAC.  The results include two-body, three-body, and quasi-two-body decay modes.  We also report measurements of polarization in modes with two vector mesons in the final state, which are also a sensitive test of SM predictions and the effect of potential non-SM contributions.

\section{Experimental Methods}\label{subsec:methods}

\subsection{Signal Extraction}

Signal decays are separated from background decays using unbinned
extended maximum-likelihood fits to distributions of kinematic 
and event-shape variables.  The primary kinematic variables used to identify
a reconstructed signal $B$ candidate are the difference $\Delta E$ between its reconstructed 
energy in the center-of-mass (CM) frame and the beam energy; and a 
beam-energy substituted mass, $m_{\rm ES} \equiv\sqrt{(s/2+{\mathbf {p}}_i
\cdot{\mathbf {p}}_B)^2/E^2_i-{\mathbf {p}}^2_B}$, where the $B$-candidate momentum
 ${\mathbf {p}}_B$ and the four-momentum of the initial $e^+e^-$ state 
$(E_i, {\mathbf {p}}_i)$ are calculated in the laboratory frame.  Event-shape variables
are used to suppress the dominant ``continuum'' $e^+e^- \to q\bar{q}$ $(q=u,d,s,c)$ 
background further, exploiting angular differences between 
the jet-like topology of continuum decays and the isotropically distributed decays 
of $B\Bbar$ events. Backgrounds from $B\Bbar$ decays into final states with charm quarks are suppressed by invariant-mass vetoes on charmonium and $D$ mesons, while backgrounds from charmless processes are rejected with selection criteria on $\Delta E$ and invariant-mass window selections and mass constraints on composite mesons in the signal decay. Particle-identification information is used to separate charged pion from charged kaon candidates in the $B^+ \to \Kzb K^+$ and $B^+ \to K^+ K^-$ decays. Angular variables are used for further signal-background discrimination and to identify helicity and polarization information in modes involving vector or tensor mesons.

\subsection{$CP$ Asymmetries}

$CP$ asymmetries in neutral $B$ decays to $CP$ eigenstates are determined from the difference in the time-dependent decay rates for $\Bzb$ and $B^0$ signal decays. The parameter $S$ describes $CP$ violation in the interference
between mixed and unmixed decays into the same final state, while $C$ describes
direct $CP$ violation in decay. If no time-dependent measurement is performed, an integrated flavor or charge asymmetry can be measured:
\begin{equation}
{\mathcal A}_{CP} = \left(N_{B^0,B^+} - N_{\Bzb,B^-}\right)/\left(N_{B^0,B^+} + N_{\Bzb,B^-}\right)
\end{equation}
A non-zero value of this asymmetry signifies the presence of direct $CP$ violation. In the charged $B$ modes, this is the only possible $CP$ measurement.\footnote{${\mathcal A}_{CP}=-C$.}

\section{Experimental Results}

\subsection{Two-Body Modes with Only Kaons and Pions in the Final State}

The $\pi\pi$ modes are important for the extraction of the angle $\alpha$,\cite{alpha} while direct $CP$ violation has been observed in the $B^0\to K^+\pi^-$ and $B^0\to\pi^+\pi^-$ modes.  In addition, several relations between branching fractions and charge asymmetries in the $B\to K\pi$ modes can be used to test SM predictions.  No significant deviations between experiment and theory is observed in the recent results, relaxing the so-called ``$K\pi$ Puzzle'' tension.\cite{KpiPuzzle}  \babar{} has also observed the $b\to d$ penguin-dominated modes $B^0 \to K^0\Kzb$ and $B^+ \to \Kbar K^+$, and measured the time-dependent $CP$-violating asymmetries in the former for the first time, utilizing a beam-constrained technique to vertex the signal $B$ meson in the absence of prompt charged tracks.  The $B^0 \to K^+K^-$ mode is the last mode left to be observed in this class of decays. Table~\ref{table1} summarizes the most recent \babar{} results.\cite{twobody}

\begin{table}[!thb]
\tbl{Branching fractions and $CP$ asymmetries for two-body modes with only kaons and pions in the final state.}
{\footnotesize
\begin{tabular}{@{}lccc@{}}
\hline
%{} &{} &{} &{} &{}\\[-1.5ex]
Mode & ${\mathcal B},~10^{-6}$  & ${\mathcal A}_{CP}$ or $-C$ & $S$ \\
\hline
%{} &{} &{} &{} &{}\\[-1.5ex]
$B^0\to\pi^+\pi^-$ & $5.5 \pm 0.4 \pm 0.3$ & $0.21 \pm 0.09 \pm 0.02$ & $-0.60\pm 0.11\pm 0.03$\\
$B^+\to\pi^+\pi^0$ & $5.1 \pm 0.5 \pm 0.3$ & $-0.02 \pm 0.09 \pm 0.01$ & $-$\\
$B^0\to\pi^0\pi^0$ & $1.48 \pm 0.26 \pm 0.12$ & $0.33\pm 0.36\pm 0.08$ & $-$\\
$B^0\to K^+\pi^-$  & $19.1 \pm 0.6 \pm 0.6$ & $-0.107 \pm 0.018 ^{+0.007}_{-0.004}$ & $-$\\
$B^+\to K^+\pi^0$  & $13.3 \pm 0.6 \pm 0.6$ & $0.016 \pm 0.041 \pm 0.012$ & $-$\\
$B^+\to K^0\pi^+$  & $23.9 \pm 1.1 \pm 1.0$ & $-0.029 \pm 0.039 \pm 0.010$ & $-$\\
$B^0\to K^0\pi^0$  & $10.5 \pm 0.7 \pm 0.5$ & $-0.20\pm 0.16\pm 0.03$ & $0.33\pm 0.26\pm 0.04$\\
$B^0\to K^0\Kzb$  & $1.08 \pm 0.28 \pm 0.11$ & $0.40\pm 0.41\pm 0.06$ & $-1.28^{+0.80}_{-0.73}{}^{+0.11}_{-0.16}$\\
$B^+\to \Kzb K^+$  & $1.61 \pm 0.44 \pm 0.09$ & $0.10\pm 0.26\pm 0.03$ & $-$\\
$B^0\to K^+ K^-$  & $<0.5$ ($90\%$ C.L.) & $-$ & $-$\\
\hline
\end{tabular}\label{table1} }
\vspace*{-13pt}
\end{table}

\subsection{Vector-Pseudoscalar Decays}

\babar{} reports the first observation, at the level of $7.9\sigma$ (including systematic uncertainties), of the $b\to s$ penguin-dominated decay $B^+\to\rho^+K^0$.  The branching fraction is in agreement with the SM prediction $p'_V = -p'_P$, which is a relation between amplitudes where the spectator quark is present in the vector and pseudoscalar meson, respectively. The charge asymmetry is consistent with zero and the SM expectation.  \babar{} also presents an updated upper limit on the branching fraction of the $b\to d$ penguin-dominated decay $B^+\to \Kbar^{*0} K^+$.  Using the technique described in Ref.~\cite{DeltaS}, an improved upper limit is placed on the difference between $\sin(2\beta_{\rm eff})$ and $\sin(2\beta)$ in the $B^0\to\phi K^0$ decay mode: $|\Delta S_{\phi K^0}|<0.11$. The results are summarized in Table~\ref{table2}.\cite{VP}

\begin{table}[!tb]
\tbl{Branching fractions and $CP$ asymmetries in vector-pseudoscalar modes.}
{\footnotesize
\begin{tabular}{@{}lcc@{}}
\hline
%{} &{} &{} &{} &{}\\[-1.5ex]
Mode & ${\mathcal B},~10^{-6}$  & ${\mathcal A}_{CP}$ \\
\hline
%{} &{} &{} &{} &{}\\[-1.5ex]
$B^+\to\rho^+K^0$ & $8.0 ^{+1.4}_{-1.3} \pm 0.5$ & $-0.122 \pm 0.166 \pm 0.020$\\
$B^+\to \Kbar^{*0} K^+$ & $<1.1$ ($90\%$ C.L.) & $-$\\
$B^+\to \Kbar^{*0}_0 (1430) K^+$ & $<2.2$ ($90\%$ C.L.) & $-$\\
\hline
\end{tabular}\label{table2} }
\vspace*{-13pt}
\end{table}

%\begin{figure}[!tb]
%\epsfxsize=10cm   %width of figure - will enlarge/reduce the figures
%\epsfbox{fig3.eps}
%\figurebox{2cm}{3cm}{} %to have a box alone 
%\centerline{\epsfxsize=4.1in\epsfbox{procs-fig1.eps}}   
%\caption{First 3 normalized frequencies versus release location for
%clamped simply supported beam with internal slide
%release. \label{inter}}
%\end{figure}

\subsection{Vector-Vector Modes}

As $B$ mesons are pseudoscalars, their decays to vector-vector final states are polarized.  Using the quark spin-flip argument based on the left-handed nature of the charged weak current, we expect the following hierarchy to hold in modes dominated by $b\to u$ tree and $b\to s$ penguin amplitudes:
\begin{equation}
A_0\sim1 \gg A_+\sim \frac{m_{\rm V}}{m_{B}} \gg A_- \sim \frac{m_{\rm V}^2}{m_{B}^2},
\end{equation}
where $A_{\rm h}$ is the amplitude of helicity $h$ and $m_{\rm V}$ and $m_{B}$ are the masses of the vector and $B$ mesons, respectively.\cite{Polarization}  This translates into the prediction that the fraction of longitudinal polarization $f_{\rm L}$ in the decay should be close to $1$. Other amplitudes from SM and non-SM processes could alter this expectation. The prediction has been verified in the tree-dominated $B\to\rho\rho$ and $B\to\rho\omega$ decays, with measured $f_{\rm L}$ in the range $0.8-1.0$.

\babar{} reports branching-fraction, charge-asymmetry, and polarization measurements in the $B\to\phi K^{*0}$ and $B\to\rho K^*$ decays,\cite{VV} which are thought to be dominated by $b\to s$ penguin amplitudes.  The results are summarized in Table~\ref{table3}.  The observed longitudinal-polarization fractions are approximately $f_{\rm L}\sim 0.5$ for the vector-vector modes, while the measured transverse-polarization fractions are $f_{\perp}(B^0\to\phi K^*(892)^0)=0.227\pm0.038\pm0.013$ and $f_{\perp}(B^0\to\phi K^*_2(1430)^0 = 0.045^{+0.049}_{-0.040}\pm0.013$, implying the amplitude hierarchy $|A_0|\approx |A_{+1}| \gg |A_{-1}|$. This suggests the presence of additional contributions to the total amplitude of these decays.\cite{Polarization} These can come from SM sources, such as annihilation amplitudes, electromagnetic or charming penguins, and long-distance rescattering effects; or from non-SM sources such as right-handed supersymmetric mass insertions or tensor $Z''$. Whatever their source, the additional contributions are not interfering with the nominal amplitudes to produce sizeable $CP$ asymmetries, as the measurements are consistent with the SM prediction of zero or very small $CP$ violation.  Thus, they must occupy a peculiar corner of phase space.  It is also interesting to note that these amplitudes have a different or no effect on tensor-vector modes, as $f_{\rm L}$ is close to unity for $B^0\to\phi K^*_2(1430)^0$.

\begin{table}[!tb]
\tbl{Branching fractions, $CP$ asymmetries, and fractions of longitudinal polarization in vector-vector and vector-tensor modes. Upper limits on branching fractions at $90\%$ C.L. are given for modes with less than $3\sigma$ significance, while both central values and upper limits are given for modes with significance between $3\sigma$ and $5\sigma$.}
{\footnotesize
\begin{tabular}{@{}lccc@{}}
\hline
%{} &{} &{} &{} &{}\\
Mode & ${\mathcal B},~10^{-6}$  & ${\mathcal A}_{CP}$ & $f_{\rm L}$ \\
\hline
%{} &{} &{} &{} &{}\\
$B^0\to\phi K^*(892)^0$ & $9.2 \pm 0.7 \pm 0.6$ & $-0.03 \pm 0.07 \pm 0.03$ & $0.506\pm 0.040\pm 0.015$\\
$B^0\to\phi K^*_2(1430)^0$ & $7.8 \pm 1.1 \pm 0.6$ & $-0.12 \pm 0.14 \pm 0.04$ & $0.853^{+0.061}_{-0.069}\pm 0.036$\\
$B^0\to\phi {(K\pi)}_0^{*0}$ & $5.0 \pm 0.8 \pm 0.3$ & $0.17\pm 0.15\pm 0.03$ & $-$ \\
$B^+\to\rho^0 K^{*+}$ & $<6.1$ ($90\%$ C.L.) & $-$ & $-$\\
$B^+\to\rho^+ K^{*0}$ & $9.6 \pm 1.7 \pm 1.5$ & $-0.01\pm 0.16\pm 0.02$ & $0.52\pm 0.10\pm 0.04$\\
$B^0\to\rho^- K^{*+}$ & $<12.0$ ($90\%$ C.L.) & $-$ & $-$\\
$B^0\to\rho^0 K^{*0}$ & $5.6 \pm 0.9 \pm 1.3$ & $0.09\pm 0.19\pm 0.02$ & $0.57\pm0.09\pm 0.08$\\
$B^+\to f_0 (980) K^{*+}$ & $5.2 \pm 1.2 \pm 0.5$ & $-0.34\pm 0.21\pm 0.03$ & $-$\\
$B^0\to f_0 (980) K^{*0}$ & \multicolumn{3}{l}{$2.6 \pm 0.6 \pm 0.09$ $(<4.3)$ ~~$-0.17\pm 0.28\pm 0.02$ ~~~~~~~~~~$-$}\\
\hline
\end{tabular}\label{table3} }
\vspace*{-13pt}
\end{table}

\section{Conclusion}

\babar{} reports measurements of branching fractions, $CP$ asymmetries, and polarization fractions in charmless hadronic $B$ decays. While disagreements from SM predictions are no longer apparent in two-body decays with kaons and pions, hints of previously unconsidered amplitudes from SM or non-SM contributions have been observed in vector-vector polarization measurements.  More data and further theoretical work will shed more light on this issue in the future.

\end{document}